\pdfoutput=1

\documentclass[10pt,preprint]{aastex}

\usepackage{graphicx,apjfonts,emulateapj5,onecolfloat5}
\usepackage{subfigure}
\usepackage{ulem}

\slugcomment{Draft: September 23, 2013}

\shorttitle{Radio Detection of A Candidate Neutron Star Associated 
with Galactic Center Supernova Remnant Sagittarius A East}
\shortauthors{Zhao, Morris \& Goss}

\begin{document}

\twocolumn[

\title{Radio Detection of A Candidate Neutron Star Associated with Galactic Center Supernova Remnant Sagittarius A East}

\author{Jun-Hui Zhao\altaffilmark{1}, Mark R. Morris\altaffilmark{2} \& W. M. Goss\altaffilmark{3}}
\affil{$^1$Harvard-Smithsonian Center for Astrophysics, 60
Garden Street, MS 78, Cambridge, MA 02138}
\email{jzhao@cfa.harvard.edu}
\affil{$^2$Department of Physics and Astronomy, University of California, Los Angeles, CA 90095}
\email{morris@astro.ucla.edu}
\affil{$^3$NRAO, P.O. Box O, Socorro, NM 87801, USA}
\email{mgoss@aoc.nrao.edu} 
\begin{abstract}
We report the VLA detection of the radio counterpart of the X-ray object 
referred to as the ``Cannonball'', which has been proposed to be the 
remnant neutron star resulting from the creation of the Galactic Center 
supernova remnant, Sagittarius A East.  The radio object was detected 
both in our new VLA image from observations in 2012 at 5.5 GHz and in  
archival VLA images from observations in 1987 at 4.75 GHz and in
the period from 1990 to 2002 at 8.31 GHz. The radio morphology of 
this object is characterized as a compact, partially resolved point 
source located at the northern tip of a radio ``tongue''  similar 
to the X-ray structure observed by Chandra. Behind the Cannonball, 
a radio counterpart to the X-ray plume is observed.  This object consists 
of a broad radio plume with a size of 30\arcsec$\times$15\arcsec, followed
by a linear  tail having a length of 30\arcsec.  
The compact head  and broad 
plume  sources appear to have relatively flat spectra 
($\propto\nu^\alpha$) with 
mean values of $\alpha=-0.44\pm0.08$ and $-0.10\pm0.02$, respectively; 
and the linear tail shows a steep spectrum with the mean value of
$-1.94\pm0.05$.  
The total radio luminosity integrated from these components 
is $\sim8\times10^{33}$ erg s$^{-1}$, while the emission
from the head and tongue amounts for only $\sim1.5\times10^{31}$ erg s$^{-1}$. 
Based on the images obtained from 
the two epochs' observations at 5 GHz, 
we infer the proper motion of the object: 
$\mu_\alpha = 0.001 \pm0.003$ arcsec yr$^{-1}$ and $\mu_\delta = 0.013 \pm0.003$
arcsec yr$^{-1}$. With an implied velocity of 500 km s$^{-1}$, a plausible 
model can be constructed in which a runaway neutron star surrounded by 
a pulsar wind nebula was created in the event that produced Sgr A East.
The inferred age of this object,
assuming that its origin coincides with the center of Sgr A East,  
is approximately 9000 years.
\end{abstract}
]

\section{Introduction}
Using X-ray data observed with the Chandra observatory,
\cite{park05} found a hard compact X-ray source, CXOGC J174545.5-285829, 
having  unusual X-ray characteristics  compared to most other X-ray 
sources near the Galactic center. The authors suggested that the X-ray object  
could be identified as a high-velocity neutron star, produced from the 
core-collapse supernova (SN) explosion that created the Galactic center  
supernova remnant (SNR), Sagittarius A East. The X-ray data appear to 
be consistent with previous radio studies in which Sgr A East was interpreted
simply as a supernova remnant (SNR) \citep{jone74,gree84,goss85}.
However, because of the closeness to the supermassive black hole (SMBH), 
the inferred large energy budget ($\sim10^{52 - 53}$ erg), and 
the elongated morphology and size,  alternative ideas have been proposed 
to interpret the nature of Sgr A East. For example, multiple SN 
explosions, tidal disruption of a star by the SMBH, or a hypernova 
associated with a collapsar or microquasar have been suggested 
\citep{yus87,mezg89,khok96,lee07}. Based on a comprehensive analysis 
of X-ray data, \cite{mae02} concluded that Sgr A East originated 
from a single Type II SN explosion and can be classified as a metal-rich 
``mixed morphology'' SNR formed $\sim$10,000 years ago.  The age of 
Sgr A East inferred from the various models spans a large range, 
from $1700$ yr  to $5\times10^4$ yr. The younger age is associated with a large projected 
radial  expansion velocity of $\sim2.3 \times 10^3$  km s$^{-1}$, 
as described by \cite{rock05} who simulated a SN explosion in the 
Galactic center environment.
The oldest age is based on a large input energy of $\sim4\times10^{52}$ erg
(corresponding to
the simultaneous explosion of $\sim40$ SNe) required to expand into a
giant molecular cloud of density $\sim10^4$ cm$^{-3}$,  
as indicated in the observations of dust and molecular gas 
\citep{mezg89,mezg96};
an expansion velocity of only about $70-80$ km s$^{-1}$ 
is inferred for this scenario.

\section{Observations \& Data Reduction}

\subsection{5 GHz data}
{\bf The 2012 epoch} data were obtained from our most recent observations 
of Sgr A that were carried out at 5 GHz  in the spring (March 29 and April 22) 
and summer (July 24 and 27) of 2012 using the Jansky Very 
Large Array (VLA) in the C and B array configurations,
respectively, with total bandwidth of 2 GHz. The UT dates were
used throughout the paper.
The observations were carried out in  spectral line mode with 
a channel width of 2 MHz for each of 1024 channels so that the 
frequency-dependent effects in the broadband synthesis imaging 
can be determined and addressed. The final images have been corrected
for frequency-dependent effects. The detailed process of data reduction 
(data editing, calibration and imaging) was performed using the software  
package of the Common Astronomy Software Applications (CASA) of the 
National Radio Astronomy Observatory (NRAO). Details will 
be provided in a subsequent paper \citep{zhao13}. The final image 
was constructed using multple-scale (MS) and multiple-frequency-synthesis 
(MFS) techniques with robustness weighting parameter of $-2$, achieving an rms
noise level of 10 $\mu$Jy beam$^{-1}$ at the phase center with no emission
(as reported in Table 1). At the Cannonball, the rms noise in the primary 
beam (PB) corrected image is 11.4 $\mu$Jy beam$^{-1}$.  The resulting 
synthesized full-width-half-maximum (FWHM) beam is 
1.6\arcsec$\times$0.6\arcsec~(PA=12\arcdeg), equivalent to 
that from uniform weighting. 

The B-array data observed on July 24 and 27, 2012 were used to show  
the structure  of compact radio emission associated with
the Cannonball. The MS-MFS imaging process was carried out in 
CASA by fitting the 2-GHz spectra with a Taylor expansion
for the first two terms in order to determine the Stokes I 
and  spectral index from 4.4 to 6.4 GHz.  The intensity tt0$-$image  
at the reference frequency of 5.5 GHz was derived from the zero-order 
of the Taylor approximation while the spectral index $\alpha-$image 
($\alpha$ for $\nu^\alpha$) was  derived from the second term (tt1)
in the first-order of  the Taylor approximation.  An rms noise level of  
15 $\mu$Jy beam$^{-1}$ is inferred for the radio intensity image from 
B-array data with a FWHM beam of 1.6\arcsec$\times$0.6\arcsec (10\arcdeg). 
At the Cannonball, the rms noise increases to 17 $\mu$Jy beam$^{-1}$  
in the primary-beam-corrected image.  

We note that the $\alpha-$images are
subject to large fluctuations propagated from the residual sidelobes 
in the {\rm tt1}$-$image due to imperfect uv-coverage.
For the  sources with a large S/N ($\geq 100$) within the half-power
beamwidth (HPBW) of the primary beam, the mean value of
spectral index from the $\alpha-$image is reliable 
(Rau V. Urvashi 2013, personal communication).
This statement appears to be consistent with our $\alpha-$images;
towards the regions with S/N in the range from a few hundreds to
a few thousands, such as the HII mini-spiral arms of Sgr A West and 
the HII regions (A -D) located east of Sgr A East, flat and/or 
gently-rising continuum spectra with less significant
fluctuations are in good agreement with the previous spectral 
image trends derived from 6 and 20 cm VLA data \citep{pedl89}. However, 
the Cannonball components, with typical S/N $\sim$ 50, exhibit significant 
fluctuations in the $\alpha$-image. Therefore, we only report
the mean values of spectral index and cite errors from the standard 
deviation of the mean on the Cannonball sources (Table 2).

\begin{figure*}[]
\centering
\includegraphics[angle=0,width=150mm]{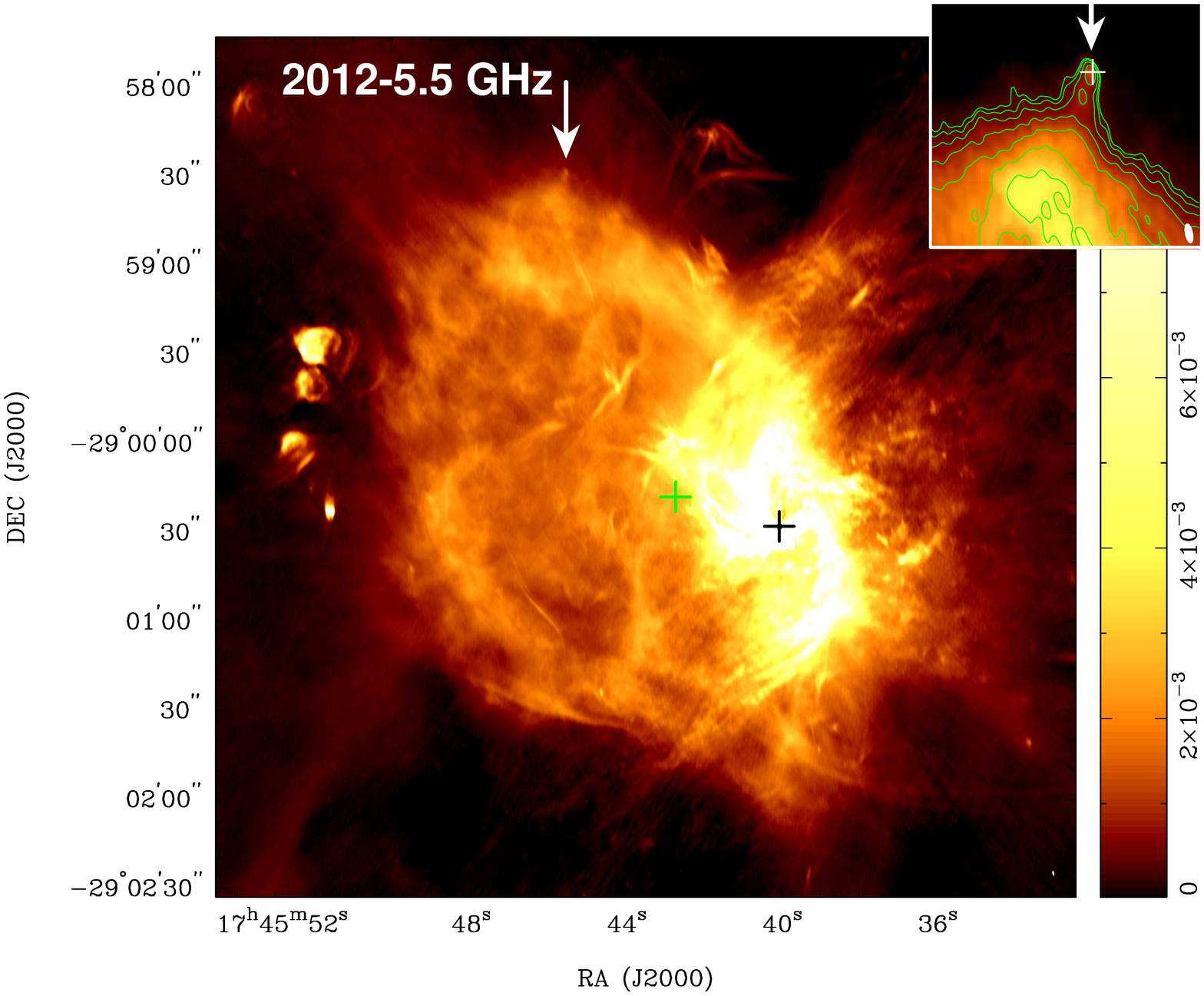} \\
\includegraphics[angle=0,width=90mm] {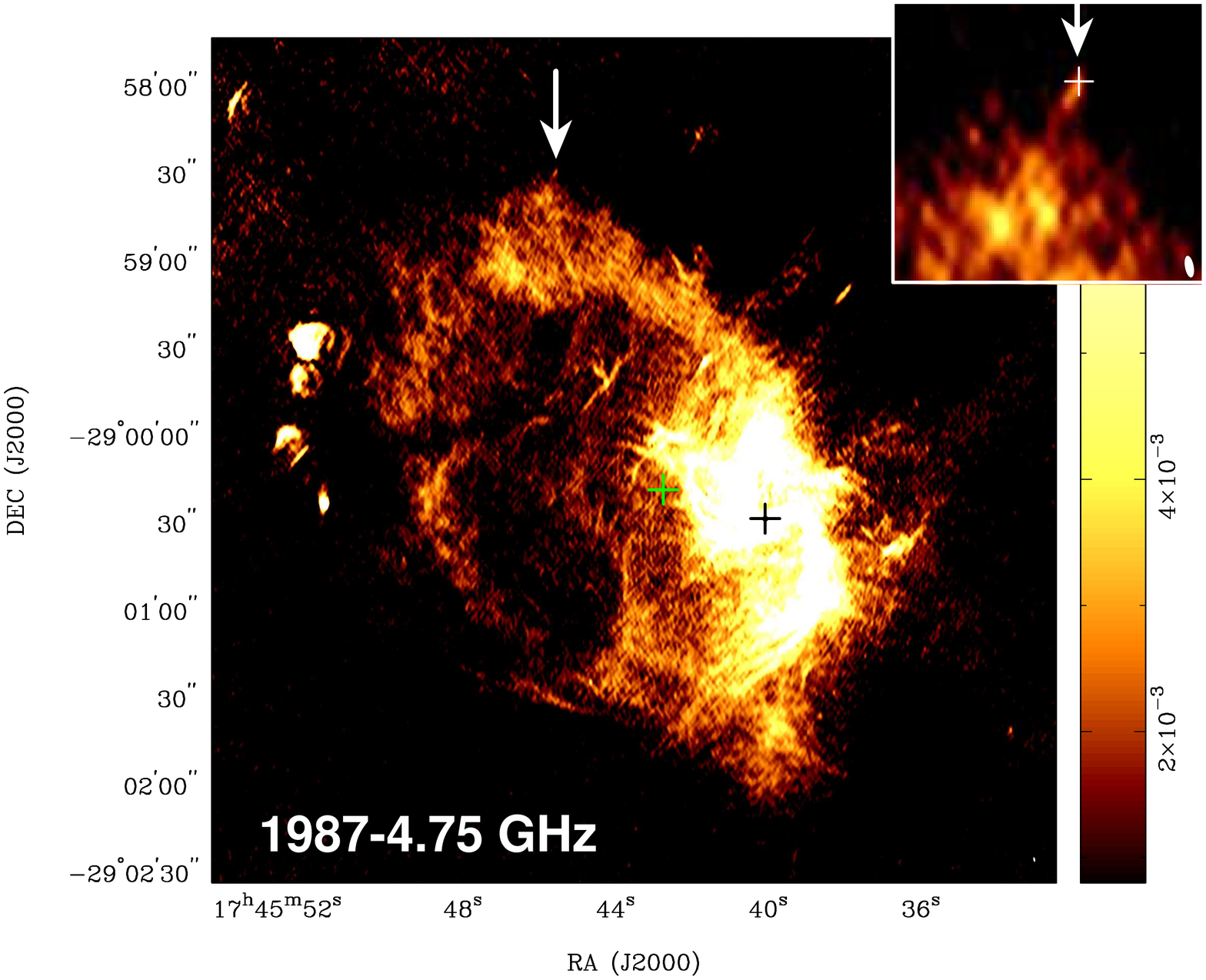}
\includegraphics[angle=0,width=90mm] {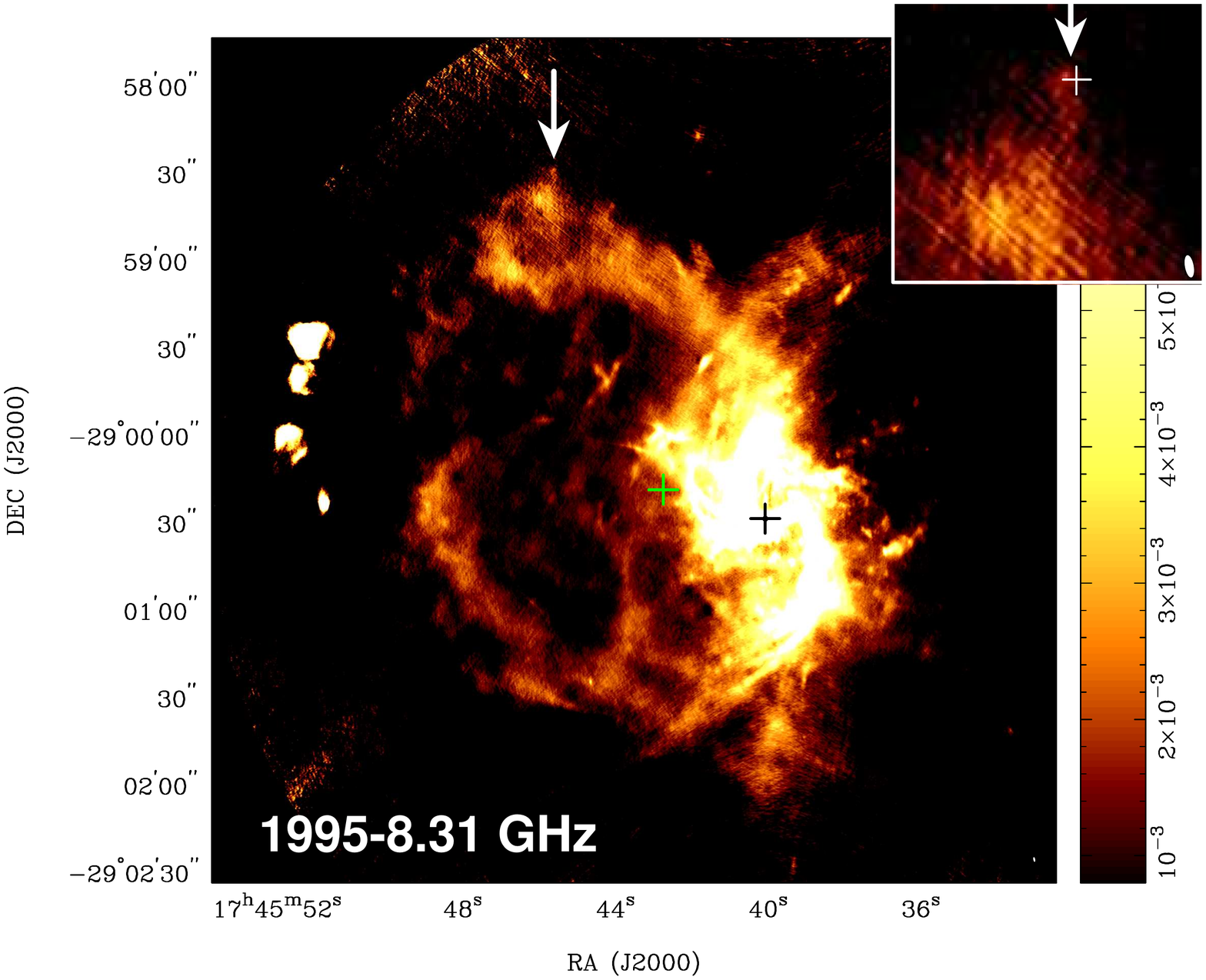}
\caption{The VLA images of Sgr A East, Top: 2012 epoch image at 5.50 GHz
constructed with the B and C array data, bottom-right: 1995 image at 8.31 GHz, 
bottom-left: image made from 1987 data  at 4.75 GHz. Sgr A* is marked as a 
black dot with a superimposed cross located at the center of Sgr A West.
The green cross marks the pointing center of the 2012 observations at 5.50 GHz,
which is close to the geometrical center of the Sgr A East shell. The radio 
counterpart of the X-ray object ``Cannonball'' was  detected in all the images,
as indicated by a white arrow. The inset at the top right of each image shows a 
22\arcsec$\times$20\arcsec\/region surrounding the Cannonball. The color wedge 
to the right indicates the radio intensity scale in units of Jy beam$^{-1}$. 
Note that the 4.75 GHz image is deconvolved to the common resolution
(1.6\arcsec$\times$0.6\arcsec, PA=10\arcdeg) which is smaller than the
synthesized beam of the 4.75 GHz data.
The FWHM beam (convolved to 1.6\arcsec$\times$0.6\arcsec, PA=10\arcdeg) 
is marked at the bottom-right corner. All these images have been corrected for 
primary beam attenuation. 
}
\label{fig:sgraeast6cm}
\end{figure*}

{\bf The 1987 epoch} data used in this study are from the VLA observations
of Sgr A at three reference frequencies, 4.585, 4.815, and 4.915 GHz, in the 
B array configuration in a spectral line mode with channel width of 3.125 MHz 
for each of 15 channels. We calibrated the data in AIPS following the VLA standard 
calibration procesure for a spectral line data set. The radio source 3C 84 was 
used for calibration of the bandpass, 3C286 for the flux density scale 
and NRAO 530 for the complex gains. The coordinates of equinox B1950 in 
the archived data were converted to J2000 using the AIPS task UVFIX. 
The spectral data were further reduced and binned to 3 channels in CASA 
for imaging with the MS-MFS technique; thus, with a channel width of 
15.6 MHz, the intensity loss due to the effect of the 
delay beam is negligible. An rms noise level of 130 $\mu$Jy beam$^{-1}$
was achieved in the final image. At the Cannonball, the rms noise increases
to 150 $\mu$Jy beam$^{-1}$ in the primary beam corrected image.

\subsection{8.3 GHz data}
An image was made from the VLA data observed at 8.31 GHz in BnA, CnB and D 
configurations in nine observations during the period between 1990 July 2 
and 2002 May 22. All the observations were carried out in spectral line 
mode with a channel width of 0.3906 MHz for 31 channels. The data were 
calibrated in AIPS (see \cite{zhao09} for the details of data reduction). 
The mean observing epoch of the 8.31 GHz data is 1995.9, which is hereafter 
referred as {\bf the 1995 epoch}. The image was made in CASA using robust=1 
weighting with the MFS technique at a center frequency of 8.31 GHz; the 
synthesized image was further cleaned with the MS technique. The rms of 
35$\mu$Jy beam$^{-1}$ was achieved with a FWHM synthesized beam of 
0.7\arcsec$\times$0.5\arcsec~(PA=69\arcdeg). At the Cannonball, the rms 
is 52 $\mu$Jy beam$^{-1}$ in the primary-beam-corrected image. The image 
used in this paper was convolved to a beam of 1.6\arcsec$\times$0.6\arcsec~(PA=10\arcdeg) 
to match the beam of the 5.5 GHz image in order to facilitate comparisons 
between the images from different frequency bands and different time epochs. 
The parameters of the final images used in this paper are summarized in Table 1.

\begin{figure*}[t]
\centering
\includegraphics[angle=0,width=150mm]{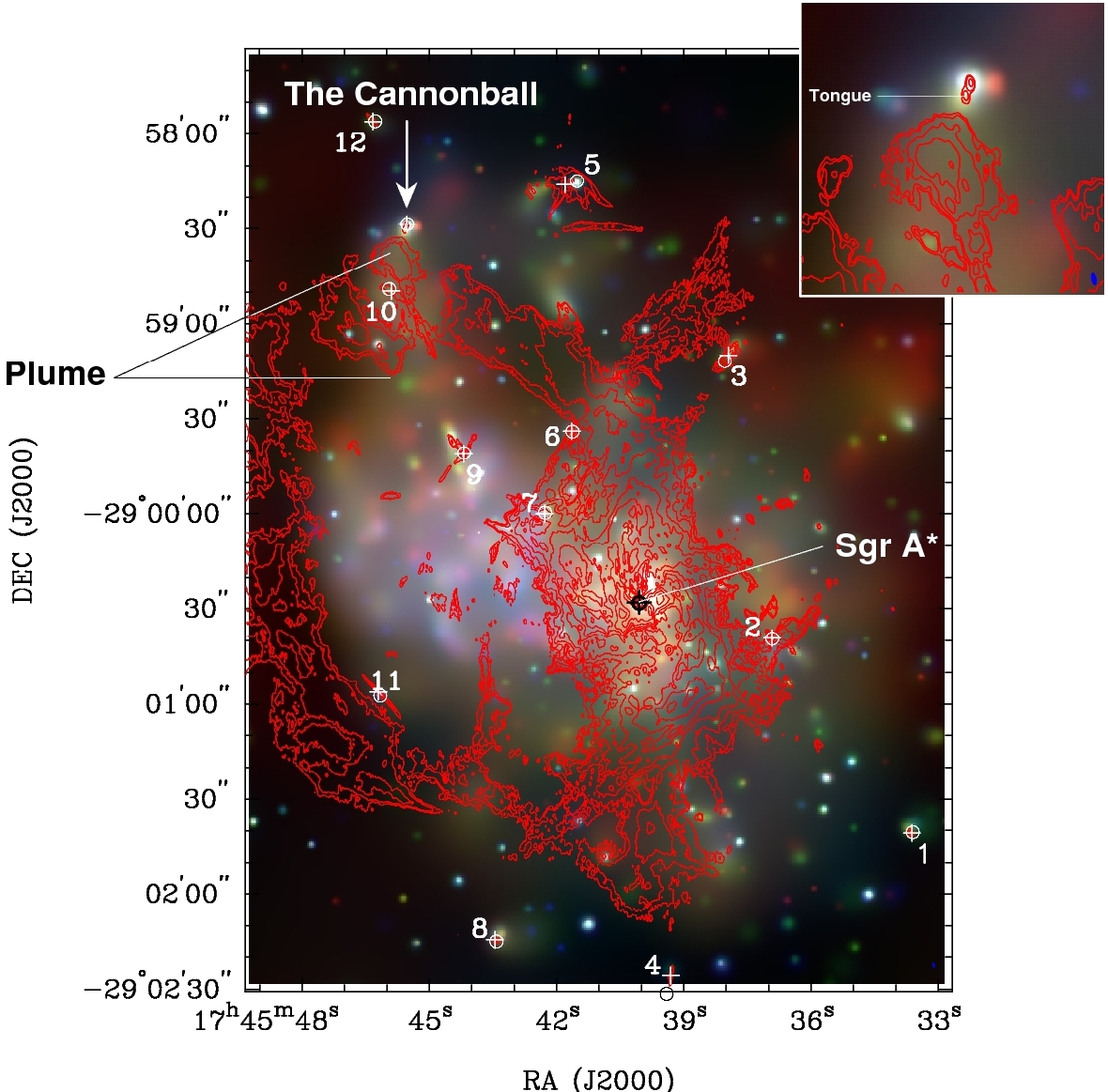}
\caption{The VLA B-array image at 5 GHz (red contours) is overlaid on
the Chandra three-color image of the X-ray object ``Cannonball'' 
(red represents the 1.5--4.5 keV band photons, green represents 
4.5--6.0 keV, and blue represents the 6.0--8.0 keV band) \citep{park05}.
The radio counterpart of the Cannonball is denoted by the white arrow. 
The X-ray plume described by  \cite{park05} is indicated. 
The black plus symbol marks the radio position of Sgr A*;
the black open circle is the X-ray position of Sgr A* \citep{muno09}.
The rest of the compact and filament radio sources (crosses)
identified with  X-ray point sources from \cite{muno09} 
(open circles) were used for the alignment between the X-ray and radio frames.
The inset at the top-right shows an enlarged image of the area {\it ($\sim40$\arcsec)}
surrounding the Cannonball.
}
\label{fig:Cannonball6cmXray}
\end{figure*}

\section{Results}
\subsection{Detection of a radio counterpart of the X-ray Cannonball}
Figure 1 shows the 2012 VLA 5.5 GHz image of the radio emission from Sgr A East. 
At the tip of a radio emission plume located at the northernmost edge of 
the Sgr A East shell, a compact radio source was detected coinciding with 
the X-ray source J174545.5-285828 \citep{muno09}\footnote{The X-ray 
source was originally referred as CXOGC J174545.5-285829 
by \cite{park05} based on the position in \cite{muno03}.}.
The compact radio source corresponding to the X-ray object was also detected 
at the northern tip of the radio plume in both the
1987 and 1995 epochs at 
4.75 and 8.31 GHz.

Figure 2 shows a comparison between the radio and X-ray sources in
the central 5\arcmin$\times$3.5\arcmin~region of the Galaxy
by overlaying the VLA B-array image at 5 GHz on the three-color
image made from Chandra observations \citep{park05}. 
The alignment of coordinate frames of the radio and X-ray images 
was made using the radio counterparts of fourteen compact X-ray 
sources with positions given in the catalog of Chandra X-ray 
point sources \citep{muno09}. Except for the well-known source Sgr A*,  
the Cannonball and a dozen  additional sources have been newly identified 
with either compact or filamentary radio sources\footnote{ 
A list of radio sources (this paper) identified with X-ray point 
sources \citep{muno09} was used for the
alignment of the radio and X-ray coordinate frames (Figure 2):\\
$
\begin{array}{lcc}
Id  & Radio-RA,~ Decl.~(J2000)  & X-ray~ID                 \\
1.  &17:45:33.610,~ -29:01:40.87& {\rm CXOUGC~J}174533.5-290140\\
2.  &17:45:36.907,~ -29:00:39.28& {\rm CXOUGC~J}174536.9-290039\\
3.  &17:45:37.934,~ -28:59:10.03& {\rm CXOUGC~J}174538.0-285911\\
4.  &17:45:39.294,~ -29:02:25.85& {\rm CXOUGC~J}174539.3-290232\\
5.  &17:45:41.788,~ -28:58:16.13& {\rm CXOUGC~J}174541.5-285814\\
6.  &17:45:41.629,~ -28:59:34.00& {\rm CXOUGC~J}174541.6-285934\\
7.  &17:45:42.269,~ -28:59:59.99& {\rm CXOUGC~J}174542.2-285959\\
8.  &17:45:43.444,~ -29:02:14.48& {\rm CXOUGC~J}174543.4-290214\\
9.  &17:45:44.176,~ -28:59:41.17& {\rm CXOUGC~J}174544.1-285940\\
10. &17:45:45.883,~ -28:58:49.72& {\rm CXOUGC~J}174545.9-285849\\
11. &17:45:46.192,~ -29:00:56.24& {\rm CXOUGC~J}174546.1-290057\\
12. &17:45:46.309,~ -28:57:56.17& {\rm CXOUGC~J}174546.2-285756\\
\end{array}
$
} which will be discussed in a subsequent paper \citep{morr13}.
The mean magnitude of the positional offset between the radio and 
X-ray sources is 1.1\arcsec$\pm$0.5\arcsec, which is much greater than  the 
positional offset (0.24\arcsec) of the Cannonball between 
the radio (this paper) and the X-ray \citep{muno09}. 
The X-ray position of Sgr A* \citep{muno09} is offset
by 0.16\arcsec~from the 
radio position determined by \cite{reid04}. Thus, limited by the uncertainty 
of the X-ray position for the reference source Sgr A*, the uncertainty in 
the alignment between radio and X-ray frames (or the rms residual
errors) is about 0.2\arcsec. We note that a large fraction of the radio 
counterparts of the compact X-ray sources are radio filaments; the 
relatively large 
positional uncertainties of the linear radio sources 
result from their confinement in only one dimension, so 
they are the dominant contribution to the mean positional offset.

In general, the  radio emission
defining the Sgr A East radio shell appears to be the extended 
radio emission that lies just outside the northern, eastern, and 
southern boundaries of the X-ray emitting region of the 
mixed-morphology SNR \citep{mae02}. The western boundary appears 
to overlap with Sgr A West, the mini-spiral of the central HII 
region and the circum-nuclear disk (CND) around the  
SMBH located at Sgr~A*.  

The Cannonball is characterized in X-ray images as a high-intensity point 
source (head) located at the tip of an X-ray plume that appears to
point toward the north-northwest, showing a tail extending to the 
south-southwest (See Figure 2). In addition, a brighter X-ray extension 
(5\arcsec) located immediately southeast of the Cannonball head is 
evident (see the inset of Figure 2). A similar extended structure 
was also seen in the X-ray image of the Mouse PWN and referred to as
a ``tongue'' \citep{gaen04}. The Mouse is a prototypical PWN; 
we find that the Cannonball is comparable to the Mouse
(see below). We have adopted the nomenclature
as described by \cite{gaen04} to describe the feature of the secondary
radio peak associated with the Cannonball head.

Similar to the X-ray morphology, the radio image also shows a compact 
head slightly resolved in the north-south direction,  connected 
with a plume embedded in the strong radio emission from the northern 
edge of the Sgr A East radio shell (Figure 1a). In the B-array image 
(Figures 2 and 3a), the compact portion of the radio plume is shown with 
enhanced contrast against the extended emission from the region
near the northern boundary of Sgr A East, which is closely associated
with the X-ray plume.   

\begin{table}
\setlength{\tabcolsep}{0.7mm}
\caption{Image Log}
\begin{tabular}{lccccc}
\hline
\hline
(1) &   (2) &   (3) & (4) & (5) & (6)  \\
Epoch & $\nu_c$ & $\Delta\nu$&$\theta_{syn}$& r.m.s.&Array \\
& (GHz)&(GHz)&Maj$\times$Min (PA)& ($\mu$Jy bm$^{-1}$)&\\
\hline
2012  & 5.50 & 2.01 &1.6\arcsec$\times$0.6\arcsec\/ (12\arcdeg)& 10&B \& C\\
2012  & 5.50 & 2.01 &1.6\arcsec$\times$0.6\arcsec\/ (10\arcdeg)& 15&B\\
1995 & 8.31 & 0.0004&0.7\arcsec$\times$0.5\arcsec\/ (69\arcdeg)& 35&Ref.\\
1987  & 4.75 & 0.05 &2.0\arcsec$\times$1.0\arcsec\/ (3\arcdeg) & 130&B\\
\hline
\end{tabular} \\
\begin{tabular}{ll}
{\footnotesize (1) Mean epoch of the images.} & {\footnotesize
(2) The center frequencies used in MFS.}\\
{\footnotesize (3) The bandwidth of the data. } &{\footnotesize (4) The synthesized beams.}\\
{\footnotesize (5) The rms noise level. } &{\footnotesize (6) The array configurations.}\\
\multicolumn{2}{l}{\footnotesize Ref: The 8.31 GHz data were summarized in \cite{zhao09}.} \\ 
\end{tabular}
\end{table}

\begin{figure*}[t]
\centering
\includegraphics[angle=0,width=190mm]{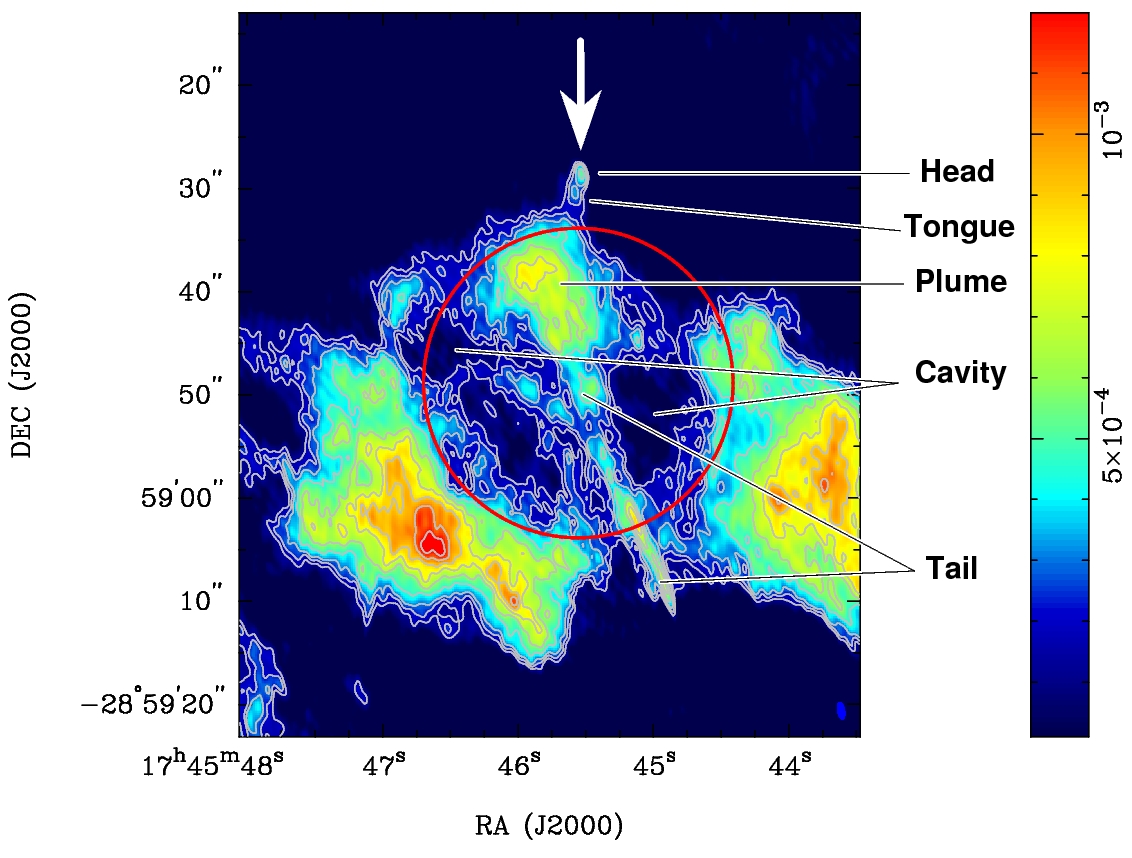}
\caption{The 2012 VLA B-array image of intensity at 5.5 GHz,
obtained from the first term Taylor expansion, shows the radio counterpart
(indicated by a white arrow) of the X-ray `Cannonball'. The color wedge
scales the radio intensity in units of Jy beam$^{-1}$ with the FWHM beam 
(1.6\arcsec$\times$0.6\arcsec, PA=10\arcdeg) shown at the bottom-right corner.
The contours are 15 $\mu$Jy beam$^{-1}\times$(10, 15, 20, 30,40, 50, 60, 70 and 80).
The red circle marks the cavity region.}
\end{figure*}

\subsection{Morphology and properties of the radio source}
Figure 3 shows an image of radio emission in the 1\arcmin\/ region 
near the Cannonball source. The morphology of the  radio emission 
can be characterized as (1) a compact point source at the northern 
tip of a radio plume, (2) a broad radio plume, and (3) a long linear 
radio emission tail, passing through a circular cavity in the 
radio emission:

{\bf The head} of the radio plume can be identified as a 
marginally resolved compact source that can be fit with a 
2D Gaussian having major and minor axes of 2\arcsec.0$\pm$0\arcsec.2
and 1\arcsec.1$\pm$0\arcsec.1 ( PA = $170\arcdeg\pm5$\arcdeg).
The result from Gaussian fitting shows an elongated structure 
($\sim$5\arcsec) in the northwest-southeast direction hereafter 
referred to as the radio ``{\bf tongue}'' followed by the 
radio plume located 5\arcsec, farther south of the compact head.
A secondary radio peak is clearly detected  $\sim$2\arcsec\/
southeast of the primary peak of the head in the 2012 image 
(see Figures 3 and 4). The X-ray image shows a similar structure, 
a few arcseconds to the southeast of the Cannonball (see Figure 2).
The flux density (excluding the secondary peak) of the Cannonball head
is 2.0$\pm0.1$ mJy with a 5.5-GHz peak of 0.51$\pm$0.03 mJy beam$^{-1}$. 
The flux density is doubled if including 
the extended emission from the tongue  (4.1$\pm0.2$ mJy).
Along with the values of the flux density and peak intensity derived at
4.75 and 8.31 GHz, we fitted the spectrum and derived the spectral index 
$\alpha=-0.4$ for the head source, which agrees with the values
of the $\alpha$ image  derived 
from the second term of Taylor expansion in the MS-MFS clean of the 
broadband data at 5.5 GHz 

{\bf The radio plume} is located $\sim$15\arcsec\/south-east of the 
compact source (head), and it can be roughly characterized by an 
ellipse (30\arcsec$\times$15\arcsec, PA=80\arcdeg), embedded in
a broad fan-shaped structure as delineated by the northern front of the radio 
emission. The northern tip of the radio plume is connected with the 
radio extension (``tongue'') of the head component. A total flux 
density of 750$\pm$150 mJy is estimated for the radio plume.
The large uncertainty stems from the uncertainties in defining 
the boundary,
as well as from the determination of the foreground/background  
emission from the Sgr A East shell. The spectral
index of this region appears to have a large error, fluctuating in 
a range from $-1$ to 1.
The mean value derived from the $\alpha$ image (Figure 3) 
is $-0.10\pm0.02$, indicating a rather flat spectrum. 

{\bf The linear tail} extends to the SSW from the plume, approximately
pointing to the geometric center of the Sgr A East shell (see the overall 
image of Sgr A East at 5.5 GHz in Figure 1). The length of this feature 
is $\sim30$\arcsec~(1.2 pc), with PA=22\arcdeg, 
slightly resolved (deconvolved size of 2\arcsec)
in the perpendicular direction. 
The  flux density from this linear feature is 150$\pm40$ mJy, peaking  
near the southern edge of the emission cavity, where a noticeable {\bf notch} 
is present (Figure 3). Along this linear feature, 
the spectral index decreases from the peak of the plume, 
toward the south, so that the steepest spectral slopes are 
found closest to the center of Sgr A East.
We infer a mean value of $\alpha=-1.94\pm0.05$ for the
tail by averaging the spectral index along it, 
showing that  the spectrum of the tail
is rather steep in comparison to other components.

{\bf The emission cavity}, with a nearly circular shape and a
diameter of 30\arcsec,  appears to be centered at 
$\alpha_{\rm J2000}=$17:45:45.6, $\delta_{J2000}=-$28:58:50. 
The linear emission tail appears to pass through the low-intensity 
``notch'' located at the southwestern edge of the cavity. 
The peak of the radio plume 
lies at the northern edge of the cavity.
We note that from the direction of the linear tail (northeastward), 
the radio head-tongue extension (nearly north) deflects  
$\sim$25\arcdeg\/ while the Cannonball travelled through the cavity.
The X-ray emission shows a similar curvature between the  head-tongue and
the plume.
 The radio properties are summarized in Table 2.

\subsection{Proper motion}
The radio counterpart of the X-ray object was detected from new
VLA observations in 2012  at 5.50 GHz and archival VLA observations at 
both 8.31 and 4.75 GHz. The largest time span between 
the observations is 25 years. In particular, the B-array observation at 4.75 GHz 
shows a significant detection ($>4\sigma$) of the compact radio source corresponding 
to the Cannonball with a similar uv-coverage as that of the B-array 
data from the most recent epoch (2012). The new 5.5 GHz observations were 
centered at the geometrical center of Sgr A East while both the 4.75 and  
8.31 GHz observations were at the position of Sgr A*.
For all epochs, the offsets 
($\Delta\alpha$ and $\Delta\delta$)
of the radio object with respect to Sgr A* 
are given in Table 2.
Therefore, the apparent proper motions due to Galactic 
rotation and  Earth's orbital motion  
cancel out.
We first made a linear regression for the positional offsets of
both $\Delta\alpha$ and $\Delta\delta$ including all the three epochs' data, 
weighted by the uncertainties ($1/\sigma^2$), with respect to
the observing-epoch time
to derive the slope using the  method and subroutines
described in \cite{bevi02}. For $\Delta\delta$, the linear fit gives
a correlation coefficient of 0.999 with a probability of no correlation $P_{nc}=$3\%,
while the fit to $\Delta\alpha$ gives a large value $P_{nc}=$90\%; the regression analysis
suggested a significant linear change in  $\Delta\delta$  but
fluctuations in $\Delta\alpha$ were consistent 
with noise. We note that 
errors for the position of the Cannonball source at the 1995 epoch  
were large due to
the S/N drop for the primary beam attenuation  and the poor image fidelity at 
the source in the 8.31 GHz image.
Also, the 1995-epoch image was made from  several datasets observed in the period
between 1990 and 2002,  so the spread in time is quite  large 
compared to that of other epoch images.
Therefore, excluding the 8.31 GHz data in our final proper motion fit, 
we calculate the slopes only from the 6-cm data in the 1987 and 2012 epochs. 
The linear-fitting plots are given in Figure 4.  
From the calculated slopes, values of $\mu_\alpha=0.001\pm0.003$ 
arcsec yr$^{-1}$ 
and $\mu_\delta=0.013\pm0.003$ arcsec yr$^{-1}$ were inferred for the 
proper motion of the Cannonball.  
 
We regridded both the 2012 and 1987 images to a common reference frame 
(the 8.3 GHz reference frame) in order to show the displacement of 
the radio object by superimposing the two images. We adopted 
the well-determined position \citep{reid04} for  Sgr A* in the 8.3 GHz 
reference frame. Figure 4 presents the 2012 image at 5.5 GHz overlaid 
on top of the 1987 image at 4.75 GHz showing that the radio object 
has moved northward from the center of Sgr A East
(Figure 4). The magnitude of the proper motion is 0.013$\pm$0.003 
arcsec yr$^{-1}$. At the Galactic center, this proper motion implies 
a transverse velocity of 500$\pm$100 km s$^{-1}$. Such a velocity 
is comparable to the space velocity of $\sim$600 km s$^{-1}$ estimated 
for the Mouse, a pulsar wind nebula located about a degree away
\citep{gaen04}, southeast of 
the Cannonball.

\begin{table*}
\setlength{\tabcolsep}{0.7mm}
\caption{Radio properties of the Cannonball object}
\begin{tabular}{lccccccc}
\hline
\hline
\\
Epoch$^{\dagger}$&$\nu_c$ &S$_{pk}^{\ddagger}$&S$_{tot}^{\ddagger}$&$\alpha$&$\theta_{maj}\times\theta_{min}$~(PA)&$\Delta\alpha^{*}$&$\Delta\delta^{*}$ \\
           & (GHz) & (mJy b$^{-1}$)&(mJy)      &        &(\arcsec$\times$\arcsec)~(\arcdeg) &($^s$)&(\arcsec) \\
\hline
Head:  \\
2012.567~&~5.50 ~&~0.51$\pm$0.03~&~2.0$\pm$0.10~&~$-0.44\pm0.08$~&~~2.0$\pm0.2\times1.1\pm0.1$(170$\pm5$) ~~&~5.488$\pm$0.001~&~119.39$\pm$0.03\\
1995.893~&~8.31 ~&~0.42$\pm$0.05~&~1.8$\pm$0.30~&~\dots       ~&~~2.9$\pm0.6\times1.9\pm0.4$(1$\pm$14)  ~~&~5.552$\pm$0.015~&~119.13$\pm$0.25\\
1987.910~&~4.75 ~&~ 0.8$\pm$0.3 ~&~2.5$\pm$0.4 ~&~\dots       ~&~~2.6$\pm0.5\times0.6\pm0.4$(159$\pm10$)~~&~5.479$\pm$0.005~&~119.08$\pm$0.07\\
Plume: \\
2012.567~&~5.50 ~&~\dots~&~750$\pm150$~&~$-0.10\pm0.02$&$\sim30\times15$ ($\sim80$)&$\sim$5.9&$\sim$110\\
Linear tail: \\
2012.567~&~5.50 ~&~\dots~&~150$\pm$40~&~$-1.94\pm0.05$&$\sim30\times$2 ($\sim$22)&$\sim$5.1&$\sim$86\\
\hline
\end{tabular}\\
\begin{tabular}{p{0.9\textwidth}}
\\
$^{\dagger}${\footnotesize The observing epochs listed here are the average of the observing dates
 at each of the frequency bands.} \\
$^{\ddagger}${\footnotesize Peak intensity ($S_{pk}$)
with the FWHM beam =1.6\arcsec$\times$0.6\arcsec (PA=10\arcdeg);
total flux density  ($S_{tot}$) from the corresponding source area. 
The flux densites of the Head do not include that of the Tongue. 
At the epoch of 2012.576, the total flux density of both
the Head and Tongue at 5.5 GHz is 4.1$\pm0.2$ mJy.}\\
$^{*}${\footnotesize The reference position is 
 Sgr A* $\alpha_{\rm J2000} = 17^h$45$^m40^s.0409$,
 $\delta_{\rm J2000} = -29$\arcdeg00\arcmin28.118\arcsec\citep{reid04}}\\
\end{tabular}
\end{table*}

\section{Discussion}
     
\subsection{Physical conditions \& nature of the Cannonball}

The transverse velocity of 500 km s$^{-1}$ inferred from the proper motion of 
the compact radio source at the head of the Cannonball appears to be consistent 
with the typical  magnitude of a random space velocity that a pulsar 
may have as a result of the
asymmetry in the SN explosion \citep{chat02,gaen06,zeig08}.
Integrating over a frequency range between 0.1 and 100 GHz,
assuming a power-law spectrum with the flux density at 5.5 GHz
and the spectral index given in Table 2 for each component,
we estimate the radio luminosities of the Head,
Plume and Tail (Table 3).
The luminosity of the radio emission of the Head and Tongue
amounts for 1.5$\times10^{31}$ erg s$^{-1}$.
The total radio luminosity of all three of these components is
$L_R\approx8\times10^{33}$ erg s$^{-1}$,
which is a typical value for PWN sources. The inferred $L_R$ appears 
to have the same order of magnitude as the X-ray luminosity ($L_X$) 
estimated from Chandra observations \citep{park05}. The flat spectrum of the 
radio emission from the plume ($\alpha=-0.10$) 
appears to fall in the typical range of $\alpha=0$ to
$-0.5$ of PWN sources \citep{gaen06}. 
Converting the X-ray photon index of $\Gamma=1.6$ \citep{park05} to $\alpha$ 
($\Gamma=1-\alpha$), the Cannonball spectral
index $\alpha_X=-0.6$  shows that the X-ray spectrum 
steepens by  $\Delta\alpha=0.5$, a canonical value expected from 
the aging of synchrotron particles \citep{wolt97}.
The radio properties of the Cannonball indicate that the emission is nonthermal 
synchrotron powered by the energy dissipation of a spin-down pulsar.  

The morphology of the radio source associated with the Cannonball shows 
a compact head with a trailing ``tongue'' connecting to a broad radio 
plume  followed by a long, linear tail (see Fig. 3). The radio structure 
of the PWN revealed for the Cannonball is quite similar to that of the 
Mouse, which has been well studied and modeled. Using the numerical 
simulations for the Mouse \citep{gaen04}, we have attempted an explanation 
of the observed radio components in the context of the shock structure 
associated with a PWN. Both the compact head with the elongated ``tongue'' 
observed in the radio and X-ray indicate the presence of a bow shock 
caused by the supersonic motion of the pulsar through the ISM. The bow 
shock confines the pulsar wind material in the shocked region. The tongue 
immediately behind the head, which is also observed in the Mouse  
\citep{gaen04}, represents the region where the particles are accelerated 
at the termination shock where the energy density of the pulsar wind 
is balanced by external pressure. The termination shock surface is elongated, 
having a significantly larger separation from the pulsar at the rear than 
in the direction of motion. As suggested in the simulations by \cite{gaen04}, 
the flow near the head of the bow shock advects the synchrotron-emitting 
particles back along the direction of motion of the pulsar, forming a 
broad cometary morphology as indicated by the radio plume. Material in 
this region generally moves supersonically. Directly behind the pulsar, 
material within the termination shock flows in a cylinder 
directed opposite to  the pulsar's velocity vector, forming a long narrow 
tail; the material in this region is subsonic. As also suggested by 
theoretical modeling, the  broad tail region originates in the strongly 
shocked material located immediately behind the shock front. A narrow, 
collimated tail is produced by material flowing along the path of the 
running pulsar. This is indeed the morphology that has been observed 
in the case of the Mouse \citep{yusb87,gaen04}. Usually, the first shocked 
zone of the broad tail is highly magnetized while the narrow tail is more weakly 
magnetized \citep{bucc02,roma05}.

One of the noticeable differences in the radio structure 
between the Cannonball and the Mouse is the abrupt change from the 
broad radio plume to the narrow head-tongue structure, 
as observed in the radio images. The  high-quality 2012  images   
of the Cannonball (Figure 3, for example)  show this transition  
well. If the radio-emitting 
material is confined by a bow-shock, the shape of the radio structure 
would be characterized by a Mach cone with a Mach angle ($\mu$); 
for a given Mach number $M$, ${\rm sin}~\mu = {1/ M}$.
For a smaller $M$, the shape of the bow-shock becomes blunter,
{\it i.e.} a larger $\mu$.
The transition in the shape of the radio structure from 
the plume to the head
might reflect a significant change in Mach number of the Cannonball 
as the pulsar reaches the edge of the SNR; 
{\it i.e.} the Cannonball could become 
highly supersonic  after departing from the SNR as the sound speed drops 
substantially in the ambient medium. Similiar changes in PWN morphology
were observed in the ``Guitar'' \citep{chat04}; these authors
argued that the morphology change reflects the random 
density inhomogeneities in the ISM; a high-density
region in the Guitar path created the broad rounded end
of the Guitar body.
 
Based on the radio images, the physical parameters can be calculated 
using the assumption that  the source components are in  
equipartition by minimizing  magnetic and synchrotron gas pressures. 
In the calculation, we assumed that both particle and magnetic-field 
filling factors are unity. The derived parameters are tabulated in Table 3.
The pressures of the synchrotron-emitting gas 
of 0.7, 0.4 and 5$\times10^{-8}$ dyn cm$^{-2}$
are inferred for the compact head, the broad plume and the linear tail, 
respectively. The magnetic field strengths range from 0.2--0.3 mG 
for the head and broad plume to 0.8 mG for the long linear tail. The 
strength of the magnetic field inferred for the long narrow tail
appears to be consistent with previous values inferred for 
the Galactic center region \citep{yus87b,morr89,morr07,croc10}. 
For a magnetic strength of 0.8 mG,  the synchrotron cooling 
time is about 20,000 years.

\begin{figure}[t]
\centering
\includegraphics[angle=0,width=88.5mm]{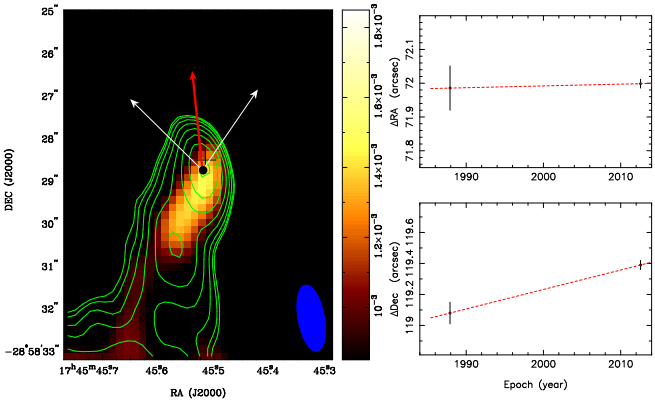}
\caption{Left: The new VLA 5.5 GHz image (contours) observed in 2012 is overlaid 
on the archival 4.75 GHz image (color) observed in 1987 to show a positional 
displacement of the radio counterpart corresponding to the X-ray object,
the Cannonball, between the two epochs. The contours are 0.12, 0.14, 0.16, 
0.20, 0.32 mJy beam$^{-1}$. The black cross marks the Chandra position of 
the X-ray source determined from the ACIS-I observations in the period from 
March 29, 2000 to July 20, 2007 \citep{muno09}.
The dot marks the VLA position of the peak emission at 5.5 GHz
(2012). 
The red arrow  anchored at the 5GHz peak indicates the proper 
motion vector ($\mu =0.013$ arcsec yr$^{-1}$ or 500 km s$^{-1}$) in P.A.
($\Theta=4$\arcdeg). The two white arrows 
indicate the  $\pm 3\sigma$  uncertainty range of
the P.A. ($\Delta\Theta_{3\sigma}=40$\arcdeg).
At the bottom-right is the FWHM beam. 
Right: The dashed lines show the motion implied by the two points separated 
in time along the horizontal axes. The vertical axis represents angular offsets 
from the common reference position.  Error bars on the positions in each 
coordinate are 1-$\sigma$. }
\label{fig:CannonballpmVect}
\end{figure}

\subsection{Trajectory and bending}

If the linear tail traces the trajectory of the Cannonball, 
then the Cannonball may have been launched at a position 
close to the geometric center of the Sgr A East 
SNR, as expected for  a runaway neutron star created by
a core collapse \citep{lai04}, 
and it appears to have travelled  along a 
straight line in PA=22\arcdeg\/ until 
it reached the boundary of the Sgr A East SNR 
where the emission cavity is located. 
Then, in this scenario, the direction of motion of the Cannonball appears to have 
changed before it departed from the cavity. The bending or 
the curvature of the 
Cannonball trajectory is the most 
remarkable feature that differs from the radio 
morphology of the Mouse. In the following analysis, we
examine a possible explanation of 
the bending or curvature of the trajectory. 
First, if the curvature  results from the
gravity of the enclosed mass within the Cannonball orbit,
the bending or deflection angle can be estimated. 
The total energy of 
the object per unit mass is given by,
\begin{equation}
 E = {1\over2} V^2 - {GM\over\/R} ,
\end{equation}
where $V$ is the velocity of the Cannonball, $R$ is the distance of the Cannonball 
to the mass center and $M$ is the enclosed mass within a radius of $R$.
Given an enclosed mass of $M=1.5\times10^7~ M_\odot$ within a radius of 5 pc 
in the central region \citep{mezg96,genz10,do13}, the total energy 
would be positive for an object traveling with a velocity of $>162$ km s$^{-1}$. 
The velocity of the Cannonball inferred from our proper motion 
measurements appears to be much greater than the escape velocity
from this region, {\it i.e.} 
$E\approx {1\over2} V^2$ for $V^2 >> 2GM/R$.
The Cannonball would then be on a 
hyperbolic orbit assuming 
Keplerian motion with respect to the center of the enclosed mass.
Deviations due to gravitational interactions with other objects
and an extended mass distribution were neglected in this approach. Then, the 
eccentricity\footnote{see http://www.braeunig.us/space/orbmech.htm} of 
the Cannonball orbit is approximately,

\begin{equation}
e \approx \left[V\over30~km~s^{-1}\right]^2\left[R\over AU\right]\left[M\over\/M_\odot\right]^{-1}.
\end{equation}
For the Cannonball, $e\approx20$ is inferred using  $R=5.4$ pc 
and $M=1.5\times10^7$ M$_\odot$.  

The turning or bending angle  is defined by the angle between the asymptotes
of a hyperbolic orbit; for the Cannonball,
\begin{equation}
\psi= {\rm arc}{\rm sin}\left(1\over e\right) \approx{1\over e}\sim 3\arcdeg,
\end{equation}
which occurs when the object passes through periapse. 
The Cannonball appears to have moved farther away from the periapse   
if it was launched near the geometric center of the Sgr A East shell;
we therefore conclude that the central enclosed mass makes an insignificant
contribution to the curvature of the Cannonball's trajectory. 
The straight path of the Cannonball, if indicated by the linear tail, 
would be consistent with this
picture.    

A possible alternative explanation for the apparent change of 
direction (aside from the trivial case that the tail 
is a coincidentally superposed, unrelated phenomenon) 
is that the Cannonball has experienced
an encounter with a massive stellar object
(or multiple encounters with a group of stars)
in the emission cavity. 
One critical problem in this explanation is that
the probability of such a close encounter is very small,
even for the high stellar density in the Galactic center region;
for the mass of a typical star (1 M$_\odot$), a close encounter  between 
the Cannonball star and a  typical field star  
at $R_{\rm enc}\approx0.03$ AU is required to deflect 
the Cannonball trajectory by 25 degrees. 
The timescale for such an encounter can be estimated by
$t_{\rm enc} = (n \sigma v_{\rm disp})^{-1} \approx 1\times 10^{13}$ yr
for stellar density $n\sim3\times10^4$ pc$^{-3}$, encounter cross section 
$\sigma\sim\pi R_{\rm enc}^2\sim 0.002$ AU$^2$,  and for a stellar velocity dispersion in the central 5 pc of $v_{\rm disp}\sim50$ km s$^{-1}$ \citep{genz10}.

However, the probability of encounter would increase if the runaway pulsar 
had been born in a binary 
system, but still remained bound to its massive 
companion as it received an impulse in the supernova
event \citep{manc06}. With  
a binary semi-major axis $a$, the encounter distance 
$R_{\rm enc} \approx a$. The encounter cross section
increases considerably after taking into account the 
gravitational focussing of orbits \citep{leon89}. 
For a binary mass of $M_{\rm b}=M_C+M_N$, 
the cross section of encountering an intruder mass of $M_{\rm i}=1$ 
M$_\odot$  
is given by \cite{moec13},
\begin{equation}
\sigma =\pi R_{\rm enc}^2\left(1+{{2G(M_{\rm b} + M_{\rm i})} \over 
{R_{\rm enc} v_{\rm disp}^2}} \right).
\end{equation}
For $M_{C}\sim$ 20 M$_\odot$ -- the mass of a high-mass companion  
such as that in J1740$-$3052 \citep{mads12},  
M$_{\rm N}\sim 1.5$ M$_\odot$ -- the mass of a neutron star 
and $R_{\rm enc}\approx a\sim10$ AU, the encounter cross section 
$\sigma$ will increase by a factor of 3$\times10^6$ as compared 
to that of the single pulsar system. Thus, the timescale for such 
an encounter is $\sim5\times10^6$ yr.
The probability of the 
observed deflection due to the gravitational bending in a binary 
encounter is greatly enhanced but is still very small compared 
to the pulsar lifetime, so this would have to be an unusually 
fortuitous occurrence.  
We therefore consider the scattering hypothesis to be unlikely, 
and do not discuss it further.

A third possibility is that, inside the Sgr A East SNR, the 
orientation of the tail is determined not only by the motion 
of the neutron star, but also by the lateral component of the 
orbital motion of the medium in which the tail is embedded: 
the SNR itself.  Once the neutron star crossed the interface 
of the SNR with the local interstellar medium (ISM), the lateral 
motion of the medium surrounding it can differ substantially, 
and thereby make it appear as if the wake, or tail, has changed 
directions.  In order to explain the 25 degree bend,
given the pulsar velocity that we infer, the lateral 
velocity difference between the two media would need to be about 
200 km s$^{-1}$, which is somewhat larger than the circular orbital velocity 
of 35 km s$^{-1}$ in this region, presenting a difficult, 
though perhaps not impossible, challenge for this hypothesis.  

An alternative explanation may be that the head-tongue association
is unrelated to the plume-tail structure. In this case, the observed 
proper motion of the neutron star is consistent with originating 
at the center of Sgr A East without any deflection.  That
 the head-tongue association does not point back 
toward the center of Sgr A East  can be explained 
by some combination of the bulk orbital motion of the SNR since its formation, 
and the differential velocity of the medium surrounding the SNR shell relative 
to the shell itself.

\begin{table}
\setlength{\tabcolsep}{0.7mm}
\caption {Physical conditions of the radio source associated with the Cannonball}
\begin{tabular}{lccc}
\hline
\hline
\\
Parameters& Head & Plume & Linear tail \\
\hline
\\
size (pc)           &~~0.08$\times$0.04~~&~~1.2$\times$0.6~~&~~1.2$\times$0.08 \\
$L_{\rm R}$ (10$^{33}$ erg s$^{-1}$)&0.01       &5 & 3.0 \\
$B_{\rm min}$ (mG)         &0.3           &0.2   & 0.8 \\
$P_{\rm min}$ (10$^{-8}$ dyn cm$^{-2}$)~~~& 0.7 &0.4& 5\\
$t_{\rm syn}$ (10$^5$ yr)        & 1     & 2   & 0.2\\
\\
\hline
\end{tabular} \\
\begin{tabular}{p{0.9\textwidth}}
{\footnotesize Row 1: the linear size;}\\
{\footnotesize Row 2: radio luminosity in 0.1 GHz $<\nu<$ 100 GHz;}\\
{\footnotesize Row 3: minimum magnetic field strength;}\\
{\footnotesize Row 4: minimum pressure of particles;}\\
{\footnotesize Row 5: synchrotron cooling time.}\\
\end{tabular}
\end{table}

\subsection{Age of the object}

The distance between the Cannonball and the estimated geometrical 
center of the Sgr A East SNR is $\sim4.7$pc (120\arcsec),
 assuming that the Sgr A East SNR is located at the same distance of 8 kpc 
as that of Sgr A* \citep{reid09}.  
Given the $\sim$500 km s$^{-1}$ lateral velocity of the Cannonball, 
the time required for traveling from the center of the SNR 
to its present position is 9000 years, neglecting any 
effect of deceleration or acceleration. If the Cannonball is a 
runaway neutron star from the core-collapse SN explosion, the 
age of Sgr A East is  the same. This agrees well with the 
SNR expansion age proposed by \cite{mezg89}. The time for a pulsar 
crossing the SNR to move into the ambient ISM is given by \cite{swal03} as,
\begin{equation}
t_{cross}= 44\times10^3~y~ \left[E_{SN}\over 10^{51}~erg\right]^{1/3}\left[n_0\over 1~cm^{-3}\right]^{-1/3}
\left[V_{PSR}\over 500~km~s^{-1}\right].
\end{equation}
For a typical pulsar in the galactic plane, the crossing time is 
$\sim$44000 yr. For the Galactic center environment, with a higher 
density ($n_0$) of the ambient medium \citep{mezg89}, a shorter crossing time 
is expected, consistent with our 9000 year estimate if 
the ambient density is about 
100 cm$^{-3}$.

\section{Conclusion}
The radio counterpart of the X-ray Cannonball was detected in our  VLA
observations in 2012 at 5.5 GHz as well as in  archival data from 
1987 at 4.75 GHz and in the period between 1990 and 2002 at 8.31 GHz. 
Based on the observed morphology and our analysis, the radio source 
can be characterized as a compact head with a trailing 
``tongue'', possibly connecting to a
broad radio plume followed by a linear tail. Such a morphology 
appears to be common for PWNe
powered by a runaway pulsar. The inferred total radio 
luminosity of $\sim8\times10^{33}$ 
erg s$^{-1}$ is consistent with the hypothesis  that the emitting relativistic electrons
in the radio source are supplied 
via the pulsar wind powered by a neutron star 
produced in the SN explosion that created the Sgr A East SNR. The inferred 
proper motion implies a transverse velocity of 500$\pm$100 km s$^{-1}$. 
If the Cannonball is associated with the Sgr A East SNR,
an age of $\sim$9000 year is inferred for both objects.

\acknowledgments
We thank Sangwook Park for providing us his three-color X-ray image.
We are grateful to R. V. Urvashi and the CASA staff for assistance 
with the data reduction. We are grateful to the referee for his/her 
valuable comments and suggestions. The Jansky Very Large 
Array (VLA) is operated by the National Radio Astronomy Observatory 
(NRAO). The NRAO is a facility of the National Science Foundation
operated under cooperative agreement by Associated Universities, Inc.
The research has made use of NASA's Astrophysics Data System.

\end{document}